\documentclass[twoside,fleqn,12pt,notebook]{article}
\usepackage{a4}
\usepackage{graphicx}
\usepackage[final]{epsfig}
\usepackage{amssymb}
\def\CellGroup{\bgroup}
\def\endCellGroup{\egroup}

\begin{document}
{\bf \LARGE Differentially rotating disks of dust\\[1cm]}
{\bf Marcus Ansorg\footnote{Friedrich-Schiller-Universit\"{a}t Jena, 
Fakult\"{a}t f\"{u}r Mathematik und Informatik, Graduiertenkolleg, 
Ernst-Abbe-Platz 4, 07743 Jena, Germany, E-mail: ansorg@mipool.uni-jena.de}
and  Reinhard Meinel\footnote{Friedrich-Schiller-Universit\"{a}t Jena, 
Theoretisch-Physikalisches Institut, Max-Wien-Platz 1, 07743 Jena, Germany, 
E-mail: meinel@tpi.uni-jena.de}\\[1cm]}
\begin{center}
{\bf Abstract}
\end{center}
\begin{quotation}\noindent
We present a three-parameter family of solutions to the stationary 
axisymmetric Einstein equations that describe differentially rotating disks of 
dust. They have been
constructed by generalizing the Neugebauer-Meinel solution of the problem of a 
rigidly rotating disk of dust. The solutions correspond to disks with
angular velocities depending monotonically on the radial 
coordinate; both decreasing and increasing behaviour is exhibited.\\
In general, the solutions are related mathematically to Jacobi's inversion 
problem and 
can be expressed in terms of Riemann theta functions. A particularly 
interesting two-parameter subfamily 
represents B\"{a}cklund transformations to appropriate seed solutions of the 
Weyl class.
\end{quotation}
KEY WORDS: Rotating bodies; disks of dust; Ernst equation; 
Jacobi's inversion problem; B\"{a}cklund transformations

\newpage
\section{INTRODUCTION}
Although many rigorous solutions to Einstein's field equations have been 
constructed, only few 
of them can be applied to physically relevant situations. In the particular 
field of stationary axisymmetric solutions so far only the Kerr black holes 
\cite{Kerr63} and the rigidly rotating disks of dust \cite{NM93,NM95,NKM96} 
are known to be physically relevant. 
On the other hand, stationary axisymmetric solutions are of special interest 
to astrophysics to 
model  equilibrium configurations like stars and galaxies.\\
The complicated structure of the Einstein equations describing the interior 
of a rotating body
gives little hope for a rigorous global solution in the near future. However, 
if one restricts oneself to 
considering only dust configurations of finite extension, the body flattens 
in an extreme manner (it becomes a disk) and
the interior equations turn into boundary conditions for the exterior vacuum 
equations. Now these equations can be expressed by a single nonlinear equation
 -- the so-called Ernst equation \cite{Ernst,Ernsta}.  For treating the Ernst 
equation, there 
are powerful analytic methods available which come from soliton theory 
\cite{Mais78}-\cite{N80b}. In particular, Korotkin \cite{Kor89,Kor93} 
and Meinel and Neugebauer \cite{MN96}, see also \cite{Kor97,MN97}, were able 
to construct a class 
of solutions containing a finite number of complex parameters and one 
arbitrary real solution 
to the axisymmetric three dimensional Laplace equation. In this paper we 
present a three-parameter subclass describing disks of 
dust revolving with a non uniform angular velocity. These solutions are 
analytic in the
sense that they belong to the class of solutions under discussion and 
therefore strictly 
satisfy the Ernst equation. On the other hand, these solutions are 
numerical solutions since 
the real-valued potential function mentioned above has been determined 
numerically in order to 
satisfy the boundary conditions. The accuracy that has been obtained was 
very high 
(generally 12 digits) such that for any practical use the solutions are 
just as good as purely analytic ones. 
Moreover, the accuracy may in principle be increased arbitrarily.\\
The paper is organized as follows. In the first chapter, the boundary 
value problem for differentially rotating 
disks of dust is introduced and the class of solutions in question is 
reviewed. The numerical methods by which we 
were able to obtain our subclass of differentially rotating disks of dust 
will 
be discussed in the first part of the second chapter. This will be followed 
by a thorough illustration of the 
parameter space of these solutions. The subsequent subchapters contain 
detailed discussions about 
particular limits. \\
In what follows, units are used in which the velocity of light as well as 
Newton's constant of gravitation are 
equal to 1.
\subsection{Metric Tensor, Ernst equation, and boundary conditions}
The metric tensor for axisymmetric stationary and asymptotically flat 
space-times reads as follows 
in Weyl-Papapetrou-coordinates $(\rho,\zeta,\varphi,t)$:
\[ds^2=e^{-2U}[e^{2k}(d\rho^2+d\zeta^2)+\rho^2d\varphi^2]
-e^{2U}(dt+a\;d\varphi)^2\,.\]
For this line element, the vacuum field equations 
are equivalent to a single complex equation -- the so-called Ernst 
equation\footnote{
The remaining function $k$ can be calculated from the Ernst potential $f$ 
by a line integration.}
   \begin{equation}
   \label{G1}
   (\Re f)\; \triangle f=({\bf \nabla} f)^2\,,
   \end{equation}
   \[\triangle=\frac{\partial^2}{\partial\rho^2}+\frac{1}{\rho}
   \frac{\partial}{\partial\rho}+\frac{\partial^2}{\partial\zeta^2},
   \qquad \nabla=\left(\frac{\partial}{\partial\rho},\frac{\partial}
   {\partial\zeta}\right),\]
   where the Ernst potential $f$ is given by
   \begin{equation}
   \label{G2}
   f=e^{2U}+\mbox{i}\,b\quad\mbox{with}\quad b_{,\zeta}=\frac{e^{4U}}
   {\rho}\,a_{,\rho},\quad
   b_{,\rho}=-\frac{e^{4U}}{\rho}\,a_{,\zeta}.
   \end{equation}
To obtain the boundary conditions for differentially rotating disks of dust, 
one has to consider
the field equations for an energy-momentum-tensor
\[T^{ik}=\epsilon u^i u^k=\sigma_p(\rho)e^{U-k}\delta(\zeta)u^i u^k,\]
where $\epsilon$ and $\sigma_p$ stand for the energy-density and the invariant 
(proper) 
surface mass-density, respectively, $\delta$ is the usual Dirac 
delta-distribution, and $u^i$ 
denotes the four-velocity of the dust material\footnote{$u^i$ 
has only $\varphi$- and $t$- components.}.  \\
Integration of the corresponding field equations from the 
lower to the upper side of the disk (with coordinate radius $\rho_0$) yields 
for $\zeta=0^+$ 
and $0\leq\rho\leq\rho_0$ the conditions (see \cite{TKl})
\begin{equation}
   \label{G3}
2\pi\sigma_p=e^{U-k}(U_{,\zeta}+\frac{1}{2}Q)
\end{equation}
\begin{equation}
   \label{G4}
e^{4U}Q^2+Q(e^{4U})_{,\zeta}+(b_{,\rho})^2=0
\end{equation}
with                                                
\begin{equation}
   \label{G5}
Q=-\rho e^{-4U}[b_{,\rho}b_{,\zeta}+(e^{2U})_{,\rho}(e^{2U})_{,\zeta}].
\end{equation}
Note that boundary condition (\ref{G4}) for the Ernst potential $f$ does not 
involve the 
surface mass-density $\sigma_p$. This condition comes from the nature of the 
material 
the disk is made of. Therefore, equation (\ref{G4}) will be referred to as the 
{\it dust-condition}.\\
The angular velocity $\Omega=u^{\varphi}/u^t$ of the disk can be calculated 
from
\begin{equation}
   \label{G6}
\Omega=\frac{Q}{a_{,\zeta}-a\;Q}\,.
\end{equation}
The following requirements resulting from symmetry conditions and asymptotical 
flatness complete our set of boundary conditions:
\begin{itemize}
\item Regularity at the rotation axis is guaranteed by 
      \[\frac{\partial f}{\partial\rho}(0,\zeta)=0.\] 
\item At infinity asymptotical flatness is realized by $U\to 0$ and $a\to 0$. 
For the 
potential $b$ this has the consequence $b\to b_\infty=\mbox{const}$. Without 
loss of generality,
this constant can be set to $0$, i.e. $f\to 1$ at infinity. 
\item Finally, we assume reflectional symmetry with respect to the plane 
$\zeta=0$, i.e. 
$f(\rho,-\zeta)=\overline{f(\rho,\zeta)}$ (with a bar denoting complex 
conjugation).

\end{itemize}      
\subsection{Solutions related to Jacobi's inversion problem}
    Meinel and Neugebauer \cite{MN96}
    showed that for an arbitrary integer $p$ the function $f$ defined by
    \begin{eqnarray}
    \nonumber
    f=\exp\left(\sum_{\nu=1}^p\int_{K_\nu}^{K^{(\nu)}}\frac{K^pdK}{W(K)}-
    v_p\right)
    \end{eqnarray}
    with 
    \begin{eqnarray}
    \nonumber
    W(K)=\sqrt{(K+\mbox{i}z)(K-\mbox{i}\bar{z})\prod_{\nu=1}^p(K-K_\nu)
    (K-\bar{K}_\nu)}
    \end{eqnarray}
    satisfies the Ernst equation\footnote{Korotkin \cite{Kor89,Kor93}, 
    see also \cite{Kor97,MN97}, found solutions to the Ernst equation 
    which are closely related to the solutions considered here.}. 
    Hereby, the $K_\nu$ are arbitrary complex parameters. 
    The variable $z=\rho+\mbox{i}\zeta$ is the 
    complex  combination of the coordinates $\rho$ and $\zeta$.                             
    The ($z$-dependent) values for the $K^{(\nu)}$ as well as the integration 
    paths on a two-sheeted 
    Riemann surface have 
    to be taken from the solution to the following Jacobian inversion problem:
   \begin{eqnarray}
    \nonumber
   \sum_{\nu=1}^p\int_{K_\nu}^{K^{(\nu)}}\frac{K^jdK}{W(K)}=v_j,
   \quad 0\leq j<p.
   \end{eqnarray}
   The potential functions $v_j\,(0\leq j\leq p)$ may be any real solutions to
   the axisymmetric Laplace equation $\triangle v_j=0$
   satisfying the recursion conditions
   \begin{eqnarray}
    \label{G9}
   \mbox{i}v_{j,z}=\frac{1}{2}v_{j-1}+zv_{j-1,z}.
   \end{eqnarray}
   These recursion conditions are automatically satisfied by the {\em ansatz}
   \begin{equation}
    \label{G10}
   v_j=\frac{1}{2\pi \mbox{i}}\int_\Sigma \frac{K^j H(K)}
   {\sqrt{(K-\zeta)^2+\rho^2}}\; dK,
   \end{equation}
   where $\Sigma$ is some curve (or even some set of curves) in the complex 
plane. 
   With (\ref{G10}), the regularity of the resulting solution at 
   $(\rho,\zeta)=
   (|\Im[K_\nu]|,\Re[K_\nu])$ is guaranteed. However, there are 
   discontinuities along the curve 
   $\Sigma'=\{(\rho,\zeta): \zeta\pm\mbox{i}\rho\in\Sigma\}$.
   Moreover, a free function $H$ defined on $\Sigma$ enters the class of 
   solutions. 
   Hence, this ansatz allows us to consider a restricted class of boundary 
   value problems\footnote{The restriction comes from the requirement that 
   the potential 
   functions $v_j$ be {\em real}.}
   in which the curve $\Sigma$ results from the shape of the boundary 
   $\Sigma'$. \\
   Ernst potentials with reflectional symmetry 
   $f(\rho,-\zeta)=\overline{f(\rho,\zeta)}$
   are characterized by the following properties:
   \begin{itemize}
   \item For each parameter $K_\nu$ there is another parameter $K_\mu$ with 
   $K_\mu=-\overline{K_\nu}$.
   \item $K\in\Sigma\,\Leftrightarrow\, \pm\bar{K}\in\Sigma$
   \item $H(\pm\bar{K})=\overline{H(K)}$
   \end{itemize}
   For differentially rotating disks, we can set 
   $\Sigma=\{K: K=\mbox{i}\rho_0x, -1\leq x\leq 1\}$. Thus we get
   \begin{eqnarray}
    \nonumber
   v_j=\rho_0^{\;j-p}\int_{-1}^1 \frac{(\mbox{i}x)^j h(x^2)}{Z_D}\;  dx,\quad 
   Z_D=\sqrt{(\mbox{i}x-\zeta/\rho_0)^2+(\rho/\rho_0)^2}\quad(\Re(Z_D)<0) .
   \end{eqnarray}
   In this expression we require the real-valued function $h$ to be analytic 
   on the interval 
   $[0,1]$. This is necessary for an analytic behaviour of the angular 
   velocity $\Omega$ for 
   all $\rho\in [0,\rho_0]$.\\ The additional requirement 
   \begin{eqnarray}
    \label{G11a}
    h(1)=0
   \end{eqnarray}
   leads to a surface mass-density $\sigma_p$ of the form
   \begin{eqnarray}
    \nonumber
    \sigma_p(\rho)=\psi(\rho)\sqrt{\rho_0^2-\rho^2}\quad\mbox{(with $\psi$ 
    analytic on $[0,\rho_0]$)   }
    \end{eqnarray}
   and therefore ensures vanishing $\sigma_p$ at the rim of the disk. The 
   resulting 
   Ernst potential $f$ depends on the normalized coordinates 
   $(\rho/\rho_0,\zeta/\rho_0)$, on the parameters $X_\nu=K_\nu/\rho_0$
   and functionally on $h$.\\
   One obtains the special case of rigid rotation \cite{NM95} when the 
   following equations are all satisfied:
   \begin{itemize}
   \item \[p=2\]
   \item \[ X_1^2=-1+\frac{\mbox{i}}{\mu},\quad X_2=-\overline{X_1}\]
   \item  \begin{equation}
         \label{G13}
         h(x^2)=\mu\frac{\mbox{arsinh}[\mu(1-x^2)]}{\pi
\sqrt{1+\mu^2(1-x^2)^2}}.
      \end{equation}
   \end{itemize}
   The parameter $\mu$ is related to the angular velocity
   \begin{eqnarray}
   \nonumber
   \mu=2\Omega^2\rho_0^2e^{-2V_0} ,\quad V_0=U(\rho=0,\zeta=0)
   \end{eqnarray}
   and runs on the interval $(0,\mu_0)$ with $\mu_0=4.62966184\ldots$ 
For $\mu\ll 1$
   one obtains the Newtonian limit of the Maclaurin disk. On the other hand,
   $\mu\rightarrow \mu_0$ and $\rho_0\to 0$ yields the ultrarelativistic limit 
of the
   extreme Kerr black hole.\\  \\
   In this article we explore the subclass $p=2$ of the solutions 
   introduced above. It will be shown how, for a given value of the complex 
parameter 
   $X_1$ ($K_1=\rho_0 X_1$), the freedom of the choice of the function $h$ 
with the 
   property (\ref{G11a}) has been used to find a solution satisfying the 
dust-condition (\ref{G4}).
   It turns out that for each $X_1$ within a certain region (a more precise 
description 
   follows) there is a function $h$ such that the resulting Ernst potential 
can be
   interpreted as having been created by a differentially rotating disk of 
dust. The accompanying 
   surface mass-density and angular velocity may afterwards be calculated
according to equations 
   (\ref{G3}) and (\ref{G6}).
\section{DIFFERENTIALLY ROTATING DISKS}
\subsection{The numerical scheme}
As already mentioned above, in this paper we consider the class of solutions 
introduced in 
the previous chapter for the particular case $p=2$. Here we can prescribe
\begin{enumerate}
\item the coordinate radius $\rho_0$
\item the complex parameter $X_1$ with $\Re(X_1)\leq 0$ and 
$\Im(X_1)\leq 0$ (without loss of generality).
Then $K_1$ and $K_2$ follow from $K_1=\rho_0 X_1,\, 
K_2=-\overline{K_1}.$
\item  a real-valued function $h:[0,1]\longrightarrow\mathbb{R}$ which is 
       analytic everywhere in $[0,1]$ (i.e. in particular at the boundaries 
of the interval) 
       and vanishes at the upper boundary: $h(1)=0$.
\end{enumerate}
For such a choice all the requirements stated in chapter 1.1 are satisfied 
except the 
dust-condition (\ref{G4})\footnote{\label{reg} Additionally one has to ensure 
that the surface 
mass-density
$\sigma_p$ given by equation (\ref{G3}) is positive and finite within 
$[0,\rho_0]$. 
Furthermore, the global regularity of the
Ernst potential has to be checked. Fortunately, our solutions possess these 
properties.}.
Now, this condition yields a complicated nonlinear integral equation for 
$h$:
\begin{eqnarray}
   \label{G14a}
\lefteqn{D\left(x^2=\frac{\rho^2}{\rho_0^2}\;;X_1;h\right)
           :=\rho_0^2\left[e^{4U}Q^2+Q(e^{4U})_{,\zeta}+(b_{,\rho})^2\right]
\doteq 0}\\&& 
\nonumber\left[\begin{array}{l}0\leq\rho\leq\rho_0,\quad\zeta=0^+,\quad
                  f=e^{2U}+\mbox{i}b=f(\rho/\rho_0,\zeta/\rho_0;X_1;h),\\[1mm]
                  Q=-\rho e^{-4U}[b_{,\rho}b_{,\zeta}
                  +(e^{2U})_{,\rho}(e^{2U})_{,\zeta}]
               \end{array}\right]
\end{eqnarray}
Note that the resulting function $h$ depends parametrically on $X_1$ (but not 
on $\rho_0$).\\
With expansions of the functions $h$ and $D$ in Chebyshev-polynomials (this 
can be done since
both of them are analytic in $[0,1]$) we try to  discretize
equation (\ref{G14a}):
\begin{itemize}
\item 
\[h(x^2)\approx\sum_{j=1}^N h_j T_{j-1}(2x^2-1)
                         -\frac{1}{2}h_1,\quad T_j(\tau)=
\cos[j\arccos(\tau)] \]
\[    h(1)\doteq 0\,\Rightarrow\,h_1=-2\sum_{j=2}^N h_j\]
\item 
\[  D(x^2;X_1;h)\approx\sum_{j=1}^{N-1} D_j(X_1;h_k) T_{j-1}(2x^2-1)
       -\frac{1}{2}D_1(X_1;h_k)\]
\end{itemize}
In this manner, the nonlinear integral equation (\ref{G14a}) is approximated 
by a finite 
system of nonlinear equations 
       \begin{equation}
       \label{G18}
       D_j(X_1;h_k)=0\quad(1\leq j<N,\quad 2\leq k\leq N).
      \end{equation}
The system (\ref{G18}) has been solved numerically by a Newton-Raphson-method. 
For this technique 
one needs a good initial guess for the solution. Fortunately, the values $h_k$ 
are given exactly
for the rigidly rotating disks. Therefore, we start with an $X_1$ that differs 
only slightly 
from those for the rigidly rotating disks 
[say $X_1^2=(-1+\varepsilon)+\mbox{i}/\mu$] and take
the $h_k$'s for the rigidly rotating disks as initial values. The newly 
calculated nearby solution
serves then as an initial guess for another solution further away from the 
rigidly rotating disks. Thus we can gradually explore the whole parameter 
region of $X_1$. \\
The numerical code written to implement this scheme produces results with 
excellent convergence. For almost 
all values $X_1$ inside the available parameter region\footnote{The 
exceptions are to be found 
in narrow stripes along the curve $\Gamma_\sigma$ (see next chapter).}
the cancellation of the terms in equation (\ref{G14a}) up to the 
12th digit and even beyond has been achieved within the whole range 
$x^2\in [0,1]$. The 
resulting Chebyshev-coefficients fall off rapidly (generally $N=20$ 
suffices to achieve the 
previously mentioned accuracy of 
12 digits) and the resulting function $h$ indeed has the desired 
smooth analytic behaviour. 
As already mentioned in footnote \ref{reg}, the accompanying surface 
mass-density turns out 
to be positive and finite and the 
Ernst potential is regular outside the disk.
\subsection{The parameter region of the solutions and examples}
In the following we discuss the solutions obtained, as a function of the 
parameter $X_1^2$.
For each value outside the hatched region in Figure 1 we have found a 
corresponding solution to the Ernst equation satisfying the dust-condition. 
What follows is a discussion of the details of 
Figure 1:
\begin{enumerate}
\item {\em Differential rotation:} For $\Re(X_1^2)=-1$ we find the rigidly 
rotating disks. On the left hand side of this line (i.e. for $\Re(X_1^2)<-1$) 
the solutions
turned out to possess 
an angular velocity $\Omega$ which increases with the radial coordinate
$\rho$. 
On the other side, for $\Re(X_1^2)>-1$, the function 
$\Omega$ decreases as $\rho$ increases.
\item {\em Ergoregions:} For values $X_1^2$ inside the area encompassed by the 
curve $\Gamma_E$, the curve $\Gamma_U$ 
and parts of the curves $\Gamma_B$ and $\Gamma_\sigma$, the corresponding 
disks 
possess an
ergoregion, i.e. a portion of the $(\rho,\zeta)$-space within which the 
function 
$e^{2U}$ 
is negative. 
\item{\em Ultrarelativistic limit:} As will be shown in chapter \ref{UL}, any 
simultaneous limit
\begin{itemize}
\item $\rho_0$ tends to $0$
\item $X_1^2$ tends to a value on $\Gamma_U$
\end{itemize}
turns out to be an ultrarelativistic limit. In the case of non vanishing 
values 
for 
$\rho^2+\zeta^2$, the resulting $f$ tends to the Ernst potential of an extreme 
Kerr black hole. 
If, on the contrary, finite values for $\sqrt{\rho^2+\zeta^2}/\rho_0$ are 
maintained, 
non asymptotically flat solutions can be obtained. 
These results are in agreement with a conjecture by Bardeen and Wagoner 
\cite{BW}.
\item{\em The Newtonian limit $|X_1^2|\to \infty$:} Here the Ernst potential 
tends to 1, i.e. 
it describes a Minkowski space. In a given neighbourhood (for large values of 
$|X_1^2|$) a 
post-Minkowskian expansion (with the first coefficient being a Newtonian 
potential) of the 
Ernst potential can be carried out. One finds that the resulting Newtonian 
coefficient is the gravitational field of a rigidly rotating Maclaurin-disk.
\item{\em The Newtonian limit $\Im(X_1)\to 0$:} For real and positive values 
of $X_1^2$ 
(hence $X_1=-X_2$, both real) again the corresponding Ernst potential is equal 
to 1. Here, the 
Newtonian coefficient of the post-Minkowskian expansion describes a disk with 
decreasing 
$\Omega(\rho)$.
\item{\em B\"{a}cklund limit:} For real and negative values of $X_1^2$ we get 
$X_1=X_2$. Then
the complicated structure of the Ernst potential $f$ simplifies considerably. 
One finds that
these solutions can be interpreted as B\"{a}cklund transforms of
appropriate seed solutions of the Weyl class. A detailed discussion on this 
subclass is given 
in chapter \ref{BL}.
\item{\em The hatched region and the curve $\Gamma_\sigma$:} Inside the 
hatched 
region no solutions 
have been found satisfying both of the requirements
\begin{itemize}
\item $f$ is regular everywhere outside the disk.
\item The function $h$ is analytic for all $x^2\in [0,1]$.
\end{itemize}
Apart from the ultrarelativistic curve $\Gamma_U$ the hatched region is 
encompassed by the
curve $\Gamma_\sigma$. Starting from $\Gamma_\Phi$ and ending at $\Gamma_U$, 
the corresponding 
Ernst potentials describe a transition from Minkowski space to the 
ultrarelativistic limit, 
just as the rigidly rotating disks and the B\"{a}cklund solutions 
(at $\Gamma_B$) do. 
Now, all solutions along $\Gamma_\sigma$ possess the property that 
the derivative of the surface 
mass-density vanishes at the rim of the disk, i.e.
$\sigma_p(\rho)=(\rho_0^2-\rho^2)^{3/2}\;\tilde{\psi}(\rho)\quad
\mbox{(with $\tilde{\psi}$ 
analytic in $[0,\rho_0]$)   }.$
Surprisingly, this physical property coincides with the failure of our 
numerical method at $\Gamma_\sigma$. Further investigations are necessary to 
clarify this 
coincidence.
\end{enumerate}
Figure 2 shows representative examples of solutions.
From Figure 3 we can get an 
impression of how ``relativistic'' our solutions are. Here, the 
physical quantity $M^2/J$ is plotted over the available parameter region 
of $X_1^2$
with the gravitational mass $M$ and the total angular momentum $J$ given 
by 
\[ M=2\int_S(T_{ab}-\frac{1}{2}Tg_{ab})n^a\xi^b dV\quad(T=g_{ab}T^{ab}),
\quad
   J=-\int_S T_{ab}n^a\eta^b dV.\]
($S$ is the spacelike hypersurface $t=constant$ with the unit future-pointing 
normal vector
$n^a$; the Killingvectors $\xi^a$ and $\eta^a$ correspond to stationarity and 
axisymmetry, 
respectively.) Note that $M$ and $J$ can also be calculated from the behaviour 
of the Ernst potential at
infinity:
\[U=-\frac{M}{r}+{\cal O}(r^{-2}),
  \quad b=-2J\frac{\cos\theta}{r^2}+{\cal O}(r^{-3})\quad(\rho=r\sin\theta,
\zeta=r\cos\theta).\]
The agreement of the two expressions yields an excellent confirmation of the 
regularity of the 
solutions. For Figure 3, not only the values $M^2/J$ have been determined 
but 
also this agreement was checked.
\subsection{Ultrarelativistic limits}
\label{UL}
In this chapter we show how ultrarelativistic limits of our solutions can be 
obtained if
simultaneously 
\begin{itemize}
\item $X_1^2$ tends to a value on $\Gamma_U$.
\item the coordinate radius $\rho_0$ tends to zero.
\end{itemize}
To this end we first consider the limit $\rho_0\to 0$ for a fixed value 
$X_1^2\notin\Gamma_U$ 
and finite values of $\rho^2+\zeta^2\neq 0$. Here, $f$ tends to $1$.
In a second step, we investigate the above simultaneous 
limit again for $\rho^2+\zeta^2\neq 0$ and find that $f$ tends to the 
Ernst potential of an extreme Kerr black hole. Finally, we indicate 
how further non 
asymptotically flat solutions can be obtained by fixing finite values of 
$\sqrt{\rho^2+\zeta^2}/\rho_0$.  
\begin{enumerate}
\item {\em The limit $\rho_0\to 0$ for $X_1^2\notin\Gamma_U$:} Similar to 
the treatment in \cite{M98}
we rewrite the expression for $f$ and Jacobi's inversion problem in the 
equivalent form\footnote{
$K^{(2)}$ is now on the other sheet of the Riemann surface.}
\begin{equation}\label{G18a}
  f=\exp\left(\int_{K^{(2)}}^{K^{(1)}}\frac{K^2dK}{W}-\tilde{v}_2\right),
\quad
  \int_{K^{(2)}}^{K^{(1)}}\frac{dK}{W}=\tilde{v}_0\,,\quad
  \int_{K^{(2)}}^{K^{(1)}}\frac{K\;dK}{W}=\tilde{v}_1\end{equation}
with \[\tilde{v}_j=v_j-w_j=v_j-\int_{K_1}^{K_2}\frac{K^j\;dK}{W}
\quad(j=0,1,2).\]
In the limit $\rho_0\to 0,\; \rho^2+\zeta^2\neq 0$ the potential 
function $v_0$ as 
well as the integral $w_0$ diverge whilst these values remain finite 
for $j>0$:
\[v_0=-\frac{1}{\rho_0\sqrt{\rho^2+\zeta^2}}\int_{-1}^1 h(x^2;X_1)\;dx+
{\cal O}(\rho_0),\quad
v_1={\cal O}(\rho_0),\quad v_2={\cal O}(\rho_0),\]
\begin{eqnarray}\nonumber\lefteqn{w_0=\frac{2}{\rho_0\sqrt{\rho^2+\zeta^2}} \;
    \Re\left[\frac{1}{X_1}K\left(\frac{X_2}{X_1}\right)\right]
+\frac{\pi \mbox{i}\zeta}{2(\rho^2+\zeta^2)^{3/2}}+{\cal O}(\rho_0)}\\ & &
\nonumber \mbox{with $K$ being the Jacobian elliptic function} \quad
            K(m)=\int_0^{\frac{\pi}{2}}\frac{d\;\theta}
{\sqrt{1-m^2\sin\theta}}
, \end{eqnarray}
\[w_1=\frac{\pi \mbox{i}}{2\sqrt{\rho^2+\zeta^2}}+{\cal O}(\rho_0),
                     \quad w_2={\cal O}(\rho_0)\]
Now, we define the curve $\Gamma_U\,(X_2=-\bar{X}_1)$:  
\begin{equation}\label{G19}
      X_1^2\in\Gamma_U\quad:\Leftrightarrow\quad C(X_1):=-\int_{-1}^1 
h(x^2;X_1)\;dx
      -2\Re\left[\frac{1}{X_1}K\left(\frac{X_2}{X_1}\right)\right]
\doteq 0
\end{equation}
For all values $X_1^2\notin\Gamma_U$, $\tilde{v}_0$ diverges as well:
\[\tilde{v}_0=\frac{C(X_1)}{\rho_0\sqrt{\rho^2+\zeta^2}}-
      \frac{\pi \mbox{i}\zeta}{2(\rho^2+\zeta^2)^{3/2}}+{\cal O}(\rho_0)
      \quad \mbox{with $C(X_1)\neq0$}\]
In the limit $\rho_0\to0$, this leads to finite solutions 
$X^{(j)}=K^{(j)}/\rho_0\,(j=1,2)$ 
of Jacobi's inversion problem and eventually to 
\[f=\exp\left(\frac{\rho_0}{\sqrt{\rho^2+\zeta^2}}
  \int_{X^{(2)}}^{X^{(1)}}\frac{X^2dX}{\sqrt{(X^2-X_1^2)(X^2-X_2^2)}}
+{\cal O}(\rho_0)\right)\to 1.\]
\item{\em The black hole limit:} The above simultaneous limit is performed 
such that 
\[\Omega_U:=\frac{C(X_1)}{2\rho_0}\] remains finite. We then get for 
$\rho_0\to 0$
 (with $\rho=r\sin\theta,\;\zeta=r\cos\theta$):
\[\tilde{v}_0=\frac{2\Omega_U}{r}-
      \frac{\pi \mbox{i}\cos\theta}{2r^2},\quad
  \tilde{v}_1=-\frac{\pi \mbox{i}}{2r},\quad \tilde{v}_2=0.\]
Again we can follow the procedure of \cite{M98}. Since $K_j\to 0\,(j=1,2)$, 
the integrals in 
Jacobi's inversion problem (\ref{G18a}) become elementary, and thus for $f$ 
one obtains
\[f=\frac{2r\Omega_U-1-\mbox{i}\cos\theta}{2r\Omega_U+1-\mbox{i}\cos\theta},\]
i.e. the ($r>0$ part of the) extreme Kerr solution with 
$J=1/(4\Omega_U^2)=M^2$. The constant 
$\Omega_U$ plays the role of the `angular velocity of the horizon'.
\item{\em The non asymptotically flat ultrarelativistic limit:} 
As shown in \cite{M98} for the rigidly rotating disks, 
our more general solutions also allow for a different, not asymptotically 
flat limit. To achieve this, one
has to consider finite values of $r/\rho_0$. A coordinate transformation
\[\tilde{r}=\frac{r}{C(X_1)},\quad \tilde{\varphi}=\varphi-\Omega_U\; t\quad
\tilde{\theta}=\theta,
\quad\tilde{t}=C(X_1)\; t\quad
\left(\mbox{hence}\quad r/\rho_0=2\tilde{r}\Omega_U\right)\]
yields a transformation to a corotating system (with angular velocity 
$\Omega_U$) combined with 
a rescaling of $r$ and $t$. In the limit $\rho_0\to 0$ and $C(X_1)\to 0$, 
the resulting 
Ernst potential $\tilde{f}$ [which is related to the
Ernst potential $f'_U$ within the above corotating system by 
$\tilde{f}=f'_U/C^2(X_1)$] 
still describes a disk and is regular everywhere 
outside the disk. However, it is not asymptotically flat. 
\end{enumerate}
\subsection{B\"{a}cklund limit}
\label{BL}
In the limit of real and negative values of $X_1^2$ we obtain purely
imaginary values for $K_1$ and $K_2$ with $K_1=K_2$. Then 
the reformulation (\ref{G18a}) of the expressions for $f$ and Jacobi's 
inversion 
problem, leads to [with the abbreviation 
$r_1 = \sqrt{(K_1-{\rm i}\bar{z})(K_1+{\rm i}z)}\,]$ :
\begin{eqnarray}
\nonumber
\lefteqn{f=\exp \left(\int_{K^{(2)}}^{K^{(1)}}\frac{dK}
{\sqrt{(K-{\rm i}\bar{z})
(K+{\rm i}z)}} - [v_2-K_1^2v_0]\right),} \\ & &
\nonumber
\int_{K^{(2)}}^{K^{(1)}}\frac{dK}{(K-K_1)\sqrt{(K-{\rm i}\bar{z})
(K+{\rm i}z)}} = v_1 + K_1 v_0 - \frac{{\rm i}\pi}{r_1},\\ & &
\nonumber
\int_{K^{(2)}}^{K^{(1)}}\frac{dK}{(K+K_1)\sqrt{(K-{\rm i}\bar{z})
(K+{\rm i}z)}} = v_1 - K_1 v_0.
\end{eqnarray}  
Applying the substitution $\lambda(K)=\sqrt{(K-{\rm i}
\bar{z})/(K+{\rm i}z)}$,
we get the following system of equations for $f$, 
$\lambda^{(1)}\lambda^{(2)}$, and $(\lambda^{(2)}-\lambda^{(1)})$ :
\begin{eqnarray}
\nonumber
\lefteqn{
f=\frac{\lambda^{(1)}\lambda^{(2)}+(\lambda^{(2)}-\lambda^{(1)})-1}
{\lambda^{(1)}\lambda^{(2)}-(\lambda^{(2)}-\lambda^{(1)})-1}\,\,
{\rm e}^{-(v_2-K_1^2v_0)},}\\&&
\nonumber
\frac{\lambda^{(1)}\lambda^{(2)}-\lambda_1(\lambda^{(2)}-\lambda^{(1)})
-\lambda_1^2}{\lambda^{(1)}\lambda^{(2)}+\lambda_1(\lambda^{(2)}
-\lambda^{(1)})
-\lambda_1^2}=-\,{\rm e}^{r_1(v_1 + K_1 v_0)},\\& &
\nonumber
\frac{\lambda^{(1)}\lambda^{(2)}-\lambda_1^*(\lambda^{(2)}-\lambda^{(1)})
-\lambda_1^{*2}}{\lambda^{(1)}\lambda^{(2)}+\lambda_1^*(\lambda^{(2)}-
\lambda^{(1)})
-\lambda_1^{*2}}={\rm e}^{r_1^*(v_1 - K_1 v_0)}\\&&
\nonumber
\qquad[\mbox{with}\quad \lambda^{(j)}=\lambda(K^{(j)}),\lambda_1=\lambda(K_1), 
\lambda_1^*=1/ \bar{\lambda}_1,  r_1^*=\bar{r}_1].\end{eqnarray}
The solution for $f$ is given by
\begin{equation}
f=f_0\,\,\frac{\left|\begin{array}{rcc}1&1&1\\-1&\lambda_1\alpha_1&
\lambda_1^*\alpha_1^*\\1&\lambda_1^2&\lambda_1^{*2}\\ \end{array}\right|}
{\left|\begin{array}{rcc}1&1&1\\1&\lambda_1\alpha_1&
\lambda_1^*\alpha_1^*\\1&\lambda_1^2&\lambda_1^{*2}\\ \end{array}\right|}
\label{BTF}
\end{equation}
where
\[f_0={\rm e}^{-(v_2-K_1^2v_0)},\quad
\alpha_1=\frac{1-\exp[r_1(v_1+K_1v_0)]}{1+\exp[r_1(v_1+K_1v_0)]}, \qquad
\alpha_1^*=\frac{1}{\bar{\alpha}_1}\;.\]
Equation (\ref{BTF}) represents a B\"acklund transformation of the real seed
solution $f_0$, see \cite{N79,N80a}. As a consequence  of (\ref{G9}),
$\alpha_1$ satisfies the
Riccati equations\footnote{It is interesting to note that the special choice 
of the integration constant
arising from (\ref{G10}) guarantees the absence of a singular behaviour of the 
final solution at
$\rho=-\Im[K_1]$.} 
\[\alpha_{1,z}=\lambda_1(\alpha_1^2-1)\frac{f_{0,z}}{2f_0}, \qquad
\alpha_{1,\bar{z}}=\frac{1}{\lambda_1}(\alpha_1^2-1)
\frac{f_{0,\bar{z}}}{2f_0}.\]
Hence, in this limit, our solutions are physically
interesting B\"acklund transforms of nontrivial seed solutions 
of the Weyl class.
\section*{ACKNOWLEDGEMENTS}
The authors would like to thank A.~Kleinw\"{a}chter and 
G.~Neugebauer for many valuable 
discussions.

\newpage
\section*{FIGURE CAPTIONS}
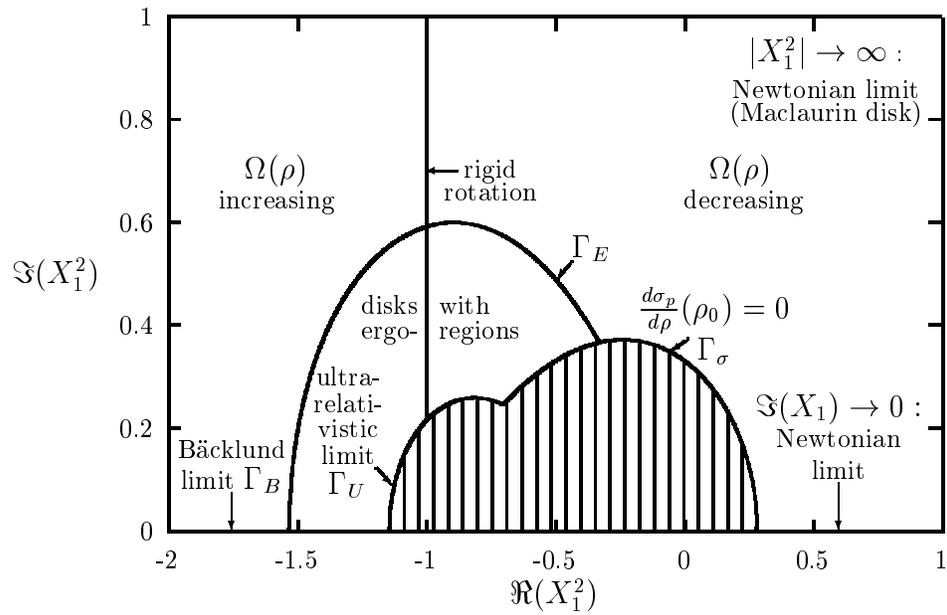
FIGURE 1: \\Outside the hatched region of the parameter space of $X_1^2$ 
regular solutions 
      have been found. Physical characteristics of solutions corresponding 
to values inside
      the allowed region are indicated.\\[1cm]
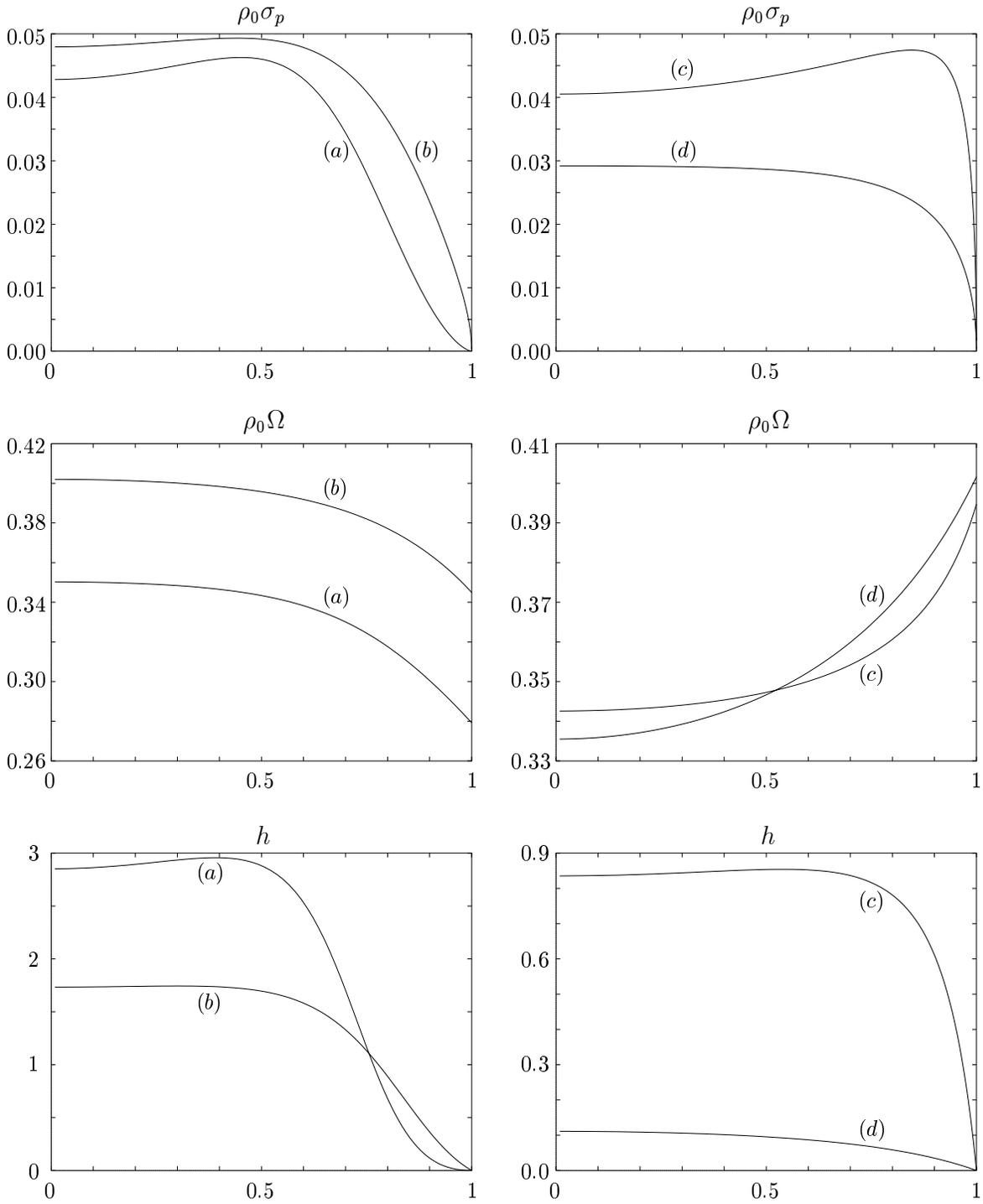
FIGURE 2: \\Examples for differentially rotating disks. The 
dimensionless 
quantities $\rho_0\sigma_p, \rho_0\Omega$ and the function $h$ 
are plotted against the 
normalized radial coordinate $\rho/\rho_0$ and $x$, respectively,  for\\[2mm]
\begin{tabular}{ll}
\hspace*{1cm}
 $(a)\quad X_1^2\approx-1/2+\mbox{i}/3$ (here $X_1^2\in\Gamma_\sigma$) 
& \hspace*{1cm}$(c)\quad X_1^2=-3/2+\mbox{i}/5$ \\[2mm] 
\hspace*{1cm}
 $(b)\quad X_1^2=-2/3+\mbox{i}/2$ & \hspace*{1cm}$(d)\quad X_1^2=-4$.\\[1cm]
\end{tabular}

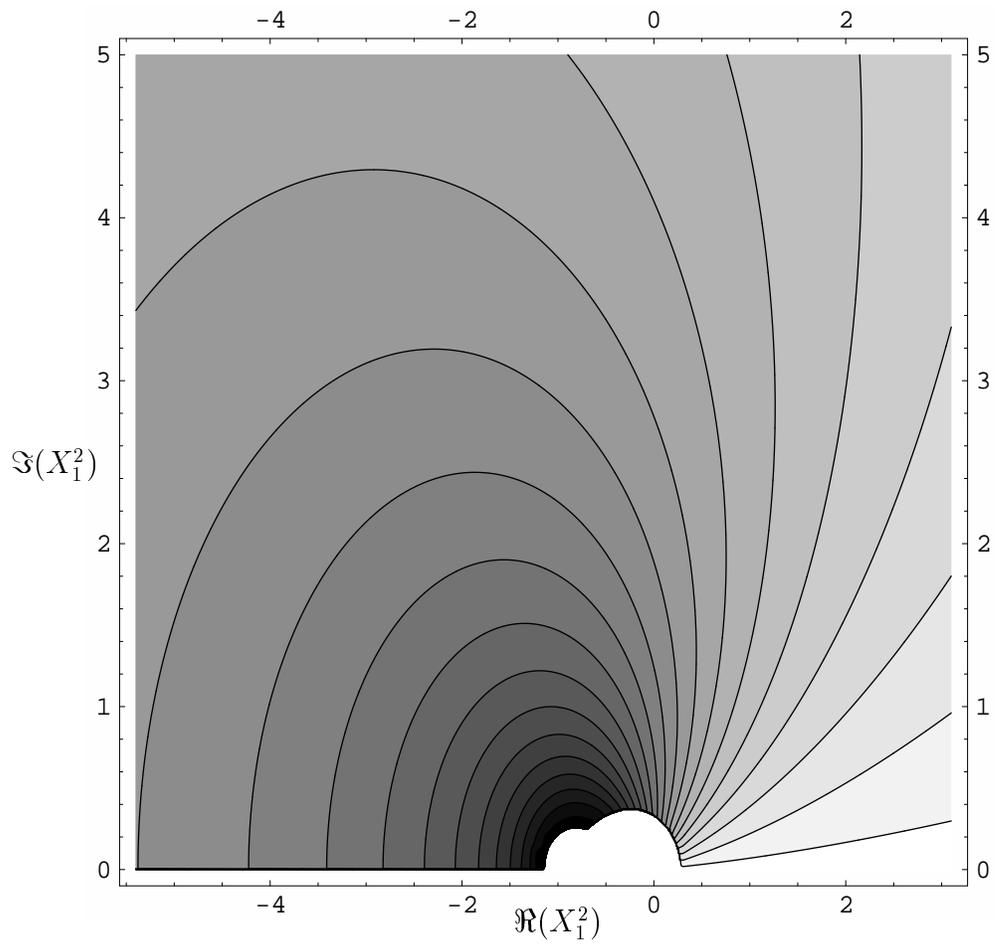
FIGURE 3: \\Contourplot of $M^2/J$. The stripes correspond to intervals of 
  length $1/20$; they are white for Newtonian disks ($M^2/J\to 0$) 
  and black for highly relativistic disks ($M^2/J\to 1$).

\end{document}